\documentclass[twocolumn,preprintnumbers,amsmath,amssymb,aps,prb,reprint,longbibliography]{revtex4-1}
\usepackage{graphicx}
\begin{document}

\title{
Collective Dynamics and Defect Generation For Wigner Crystal Ratchets }
\author{
C. Reichhardt and C. J. O. Reichhardt 
} 
\affiliation{
Theoretical Division and Center for Nonlinear Studies,
Los Alamos National Laboratory, Los Alamos, New Mexico 87545, USA
}

\date{\today}
\begin{abstract}
We consider a two-dimensional Wigner crystal coupled to a quasi-one-dimensional asymmetric potential under ac or dc driving. As a function of electron density, substrate strength, and ac amplitude, we find that the system exhibits ordered and disordered pinned and dynamical states. Ratchet effects can appear under an applied ac drive and can be associated with pronounced structural changes from a disordered state to a one-dimensional smectic-like state. We observe a pinned phase, a diode-like ratchet where motion only occurs along the easy direction of the substrate asymmetry, a plastic ratchet where motion occurs in both directions but there is only a net drift in the easy direction, and an elastic ratchet where the system forms a crystal without plastic deformation that can still undergo ratcheting. At high filling, we find that there can be a ratchet reversal in which the net drift is along the hard direction of the substrate asymmetry. For weak disorder, there is an Aubry transition to a floating phase where the ratchet effects are lost. We map out the different dynamical phases as a function of substrate strength, filling, ac amplitude, and ac frequency. The ratchet effect on strong substrates is enhanced by thermal fluctuations, but is destroyed when the fluctuations become too large. Based on our results, we suggest new ways to detect Wigner crystals as well as methods for creating new types of devices to control disordered charge flow.
\end{abstract}

\maketitle

\section{Introduction}

Wigner crystals or two-dimensional (2D) electron solids are expected
to occur at low electron densities \cite{Wigner34}
when the Coulomb energy dominates over the kinetic energy.
Early evidence for Wigner crystals was
obtained for elections on helium
\cite{Crandall71,Grimes79}, and
additional evidence has
come from solid-state systems
\cite{Doman79,Bello81,Dykman81,Andrei88,Goldman90,Jiang91,Williams91,Kopelevich07,Monceau12,Brussarski18,Hossain22}.
Recently Wigner crystal states have been found in
moir{\' e} heterostructures
\cite{Regan20,Xu20,Huang21,Li21,Matty22,Mak22,Chen23}
and dichalcogenide monolayers \cite{Smolenski21}.
Advances in
certain materials also hold promise for realizing new regimes
as well as
understanding Wigner crystals and how they interact with disorder
\cite{Falson22,Shayegan22}.
The presence of Wigner crystals
can produce
an insulating state when defects in the material 
pin the Wigner crystal;
however, at strong drives, there can be
finite depinning thresholds
and nonlinear transport
\cite{Andrei88,Goldman90,Jiang91,Williams91,Cha94,Reichhardt01,Kopelevich07,Csathy07,Brussarski18,Hossain22,Reichhardt22}.
Such depinning and nonlinear velocity-force signatures
have also been
observed in other crystalline assemblies
that exhibit
depinning,
such as vortices in type-II superconductors
\cite{Bhattacharya93,Fisher98,Reichhardt17},
charge density waves \cite{Gruner88}, frictional systems \cite{Vanossi13},
magnetic skyrmions \cite{Reichhardt22a}, and colloidal systems \cite{Bohlein12a}.

Additional evidence for Wigner crystals and their melting
has been obtained via
various types of resonance and other methods
\cite{Chen06,Knighton18,Deng19,Ma20,Kim22}.
The depinning and transport of Wigner crystals
has also been studied theoretically
and experimentally for elections on helium,
where various types of confining geometries can be created
including channels, constrictions, and
periodic one-dimensional (1D) substrates
\cite{Shirahama95,Piacente05,Araki12,Rees12,Ikegami15,GalvanMoya15,Flomenbom16,Rees16,Rees17,Moskovtsev19,Moskovtsev20,Rees20,Byeon21,Zou21}.
Evidence for Wigner crystals is generally indirect and comes from
measures such as
transport,
so other geometries
in which the crystalline or particle like nature of the Wigner crystal
could be detected would be very valuable.

Extensive studies of the ratchet effect,
where directed motion occurs for
particles that are either coupled to some form
of asymmetric substrate with an applied ac drive or
are undergoing Brownian motion on a potential that is flashing,
have been performed in
both individual and collectively interacting particle systems
\cite{Vale90,Magnasco93,Astumian94,Reimann02}.
Both flashing and rocking ratchets have been studied
in colloidal systems \cite{Rousselet94,Marquet02,Libal06},
biological systems \cite{Lau17}, active matter systems \cite{Reichhardt17a},
granular matter \cite{Farkas02}, cold atoms
\cite{MenneratRobilliard99,Lundh05}, quantum dots \cite{Platonov15},
and quantum systems \cite{Linke99,Salger09}.
In many of these systems, ratcheting
can happen on the individual particle level;
however, ratcheting motion can also occur when
collective effects become important.
Correlated ratchets have been
studied for vortices in nanostructured type-II superconductors
for 1D and 2D substrates
\cite{Lee99,Wambaugh99,Villegas03,deSouzaSilva06a,Dinis07a,Lu07,Yu07,Gillijns07,Lin11,VandeVondel11,Shklovskij14,Rouco15,Reichhardt15,Dobrovolskiy20,Lyu21},
hard sphere systems \cite{Derenyi95},
active matter systems \cite{McDermott16,Reichhardt17a}, and also
magnetic skyrmion systems \cite{Reichhardt15a,Gobel21,Ma17,Souza21,Zhang22b}.
When strong correlations are relevant,
new types of commensurate effects can occur
depending on the particle lattice
structure. When the spacing of the particle
lattice matches
the spacing of the underlying potential,
non-monotonic behavior appears in the magnitude of the directed
motion and there is an
emergence of
quasi-particles such as kinks and anti-kinks.
A particularly interesting
feature of the interacting system
is that reversals of the ratchet effect
can arise, in which
the net dc flow is in the direction opposite to the easy 
direction of the substrate 
\cite{Villegas03,deSouzaSilva06a,Lau17,Gillijns07}.
A Wigner crystal coupled with an
asymmetric substrate should provide another
example of a strongly correlated ratchet
system. If systems where Wigner crystals are expected
to form show
ratchet effects similar to those found
for superconducting vortices and other strongly correlated systems,
these ratchet effects could provide a useful
method not only to demonstrate the presence of the Wigner crystal,
but also to examine the dynamical properties of
these systems and potentially open pathways
for new applications.

There have already been some previous theoretical studies
of Wigner crystals coupled to 1D periodic modulated substrates.
Moskovtsev and Dykman considered
a 2D electron lattice on
a 1D periodic modulated surface and found
peaks and dips in
the mobility due to commensuration effects
\cite{Moskovtsev20}.
In another study, they observed
a freezing transition from
a liquid to a solid as the temperature is cooled \cite{Moskovtsev19}.
Zakharov {\it et al.} \cite{Zakharov19}
considered Wigner crystals interacting with 1D and 2D asymmetric
periodic potentials and found that an Aubry transition from
a weakly to a strongly coupled state
occurs at a critical substrate strength.
Above the Aubry transition, the system exhibits a charge diode effect.
The appearance of the diode effect suggests that the
Wigner system should also exhibit ratchet effects.

Even though
Wigner crystals
are excellent examples of strongly correlated systems that can couple
to a substrate, and
a growing number of systems are being realized that
can support Wigner crystals,
to our knowledge, Wigner crystal ratchets have yet to be studied.
One of the main differences between 
Wigner crystals and other interacting systems is the
long range of their Coulomb interactions.
For colloidal particles or vortices in type-II superconductors,
the particle-particle interaction length generally extends at most only to
a few nearest neighbors. In contrast, the interaction
extends out to all lengths for the Wigner crystals.
The long-range nature
of the interactions means that there should be transitions
even down to low charge densities
from crystalline to strongly
distorted states as a function of substrate strength, 
whereas colloidal and superconducting vortex systems
generally become disordered at lower densities since the
particles cease to interact with each other.
Additionally, for the Wigner crystals, it should be feasible to
observe both nonthermal and thermal-dominated phases.
There are several possible 1D realizations
of Wigner crystals  \cite{Schulz93,Deshpande08,Shapir19,Ziani21},
and some of the ratchet results we observe
should be visible for 1D periodic substrates
and should also occur for purely 1D systems.

Here we show that Wigner crystals coupled to a
one-dimensional asymmetric substrate
can exhibit a variety of ratchet effects for varied parameters.
These include
a pinned phase, a diode-like ratchet where the
particles move only one way during
an ac drive cycle, a plastic ratchet where the system
undergoes strong
structural rearrangements and can flow in both directions but has a
net transport in only one direction, an elastic ratchet where
the system forms a crystal that ratchets as a rigid object,
and an Aubry transition to a floating state where the
ratchet effect is lost. These different phases
can be characterized by the structural properties of the system, including the
number of defects present in the lattice
and the ratchet efficiency. 
Additionally, we find that in several of the ratchet
states, the particles
undergo distinct structural changes where the number of defects in the lattice
is strongly time dependent. These strong
structural changes likely result from the long-range nature of the
particle-particle interactions.
Our results should also be relevant
to other assemblies of Coulomb interacting particle-like objects 
coupled to asymmetric substrates,
including ions on optical traps \cite{Benassi11,Li12,Schmidt18}
and electron liquid crystal systems such as stripe and
bubble phases \cite{Fogler96,Lilly99,Reichhardt11,Friess18}.

\section{Simulation}

\begin{figure}
\includegraphics[width=\columnwidth]{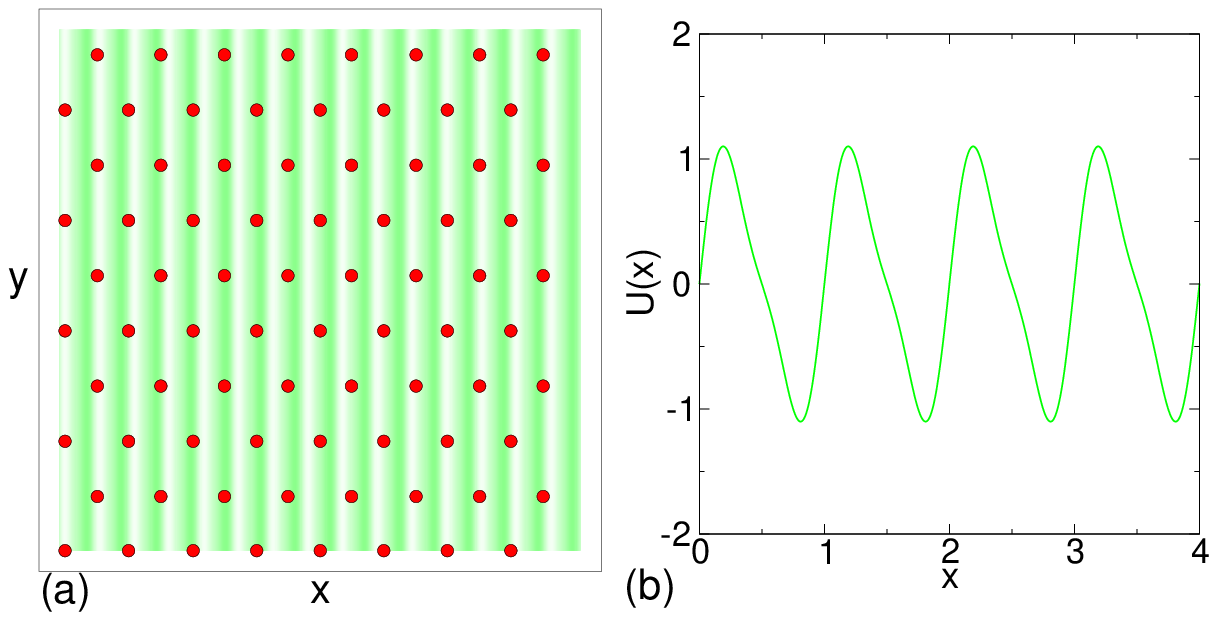}
\caption{
(a) Image of the sample showing the localized electrons (red) and the underlying
asymmetric potential (green shading). (b) The substrate potential
$U(x)$ plotted over a small range of $x$ to highlight the asymmetry.  
} 
\label{fig:1}
\end{figure}

We consider a 2D system 
of $N_e$ classical electrons
interacting with Coulomb repulsion in a sample that has
periodic boundary conditions in
the $x$ and $y$ directions.
The electrons are coupled to a
1D asymmetric substrate and the sample is of size $L \times L$ with
$L=36$. This gives an electron density of $\rho=N_e/L^2$.
The initial electron configuration is obtained via simulated 
annealing, similar to what was done in previous simulations of Wigner crystals
in the presence of random disorder
\cite{Reichhardt01,Reichhardt04,Reichhardt21,Reichhardt22}. 
The equation of motion for electron $i$ in the Wigner crystal is
\begin{equation} 
  \alpha_d {\bf v}_{i} = \sum^{N}_{j}\nabla U(r_{ij}){\bf \hat r}_{ij} +  {\bf F}_{\rm sub} +
         {\bf F}^{T}_{i}
        + {\bf F}_{D} \ .
\end{equation}
We consider the electron motion to be overdamped
with damping coefficient
$\alpha_{d}=1$.
The electron-electron interaction is obtained from
the repulsive interaction potential
$U(r_{ij}) = q/r_{ij}$,
where $q$ is the electron charge,
${\bf r}_i$ and ${\bf r}_j$ are the positions of electrons $i$ and $j$,
$r_{ij}=|{\bf r}_i-{\bf r}_j|$, and
${\bf \hat{r}}_{ij} = ({\bf r}_i - {\bf r}_j)/r_{ij}$.
Due to the long range of the interactions,
we not only cannot cut off the interactions beyond a certain length
as in superconducting vortex and magnetic skyrmion systems,
but we must also take into account all image charges from the
periodically repeated sample.
To do this, we employ the Lekner method,
a real-space version of a modified Ewald summation technique,
as in previous work \cite{Lekner91,GronbechJensen97a,Reichhardt01}.
The pinning force ${\bf F}_{\rm sub}$ is modeled as arising from
a 1D asymmetric substrate with potential
\begin{equation}
U(x) = -A_p[\sin(2\pi x/a) + 0.25\sin(4\pi x/a)] \ .
\end{equation}
This potential exerts a maximum force of
$F^{\rm hard}_p=1.5 A_p$ for particle motion in the hard or $-x$
direction, and a maximum force of
$F^{\rm easy}_p=0.752 A_p$ for particle motion in the easy or $+x$
direction.
The thermal fluctuations ${\bf F}^T$ are represented by Langevin kicks
with the properties
$\langle {\bf F}_i^{T}\rangle = 0$
and $\langle {\bf F}^T_i(t){\bf F}_j^{T}(t^\prime)\rangle = 2k_BT\delta_{ij}\delta(t-t^\prime)$.
The  applied driving force $F_D$ is either ac with
${\bf F}_{D}= F_{AC}\sin(\omega t){\bf \hat x}$ or
dc with ${\bf F}_{D}= \pm F_{DC} {\bf \hat{x}}$.
For most of this work we
use ac driving where $\omega$ is fixed at $\omega_{0} = 0.00754$. 
We measure the average velocity per electron,
$\langle V\rangle = \sum^{N_e}_i{\bf v}_i\cdot {\hat {\bf x}}$, which
we typically average over 100 to 1000 ac drive cycles.
In Fig.~\ref{fig:1}(a), we illustrate
the system with shading corresponding to
the depth of the substrate.
Figure~\ref{fig:1}(b) shows a portion of the substrate potential $U(x)$.
The total number of pinning wells
in the sample is $N_{p}$ and unless otherwise noted
we take $N_p = 17$.
The spacing between substrate minima is
$a = N_{p}/L$ and the density of the charges
is $\rho = N_{e}/L^2$.

\section{Results}

\begin{figure}
\includegraphics[width=\columnwidth]{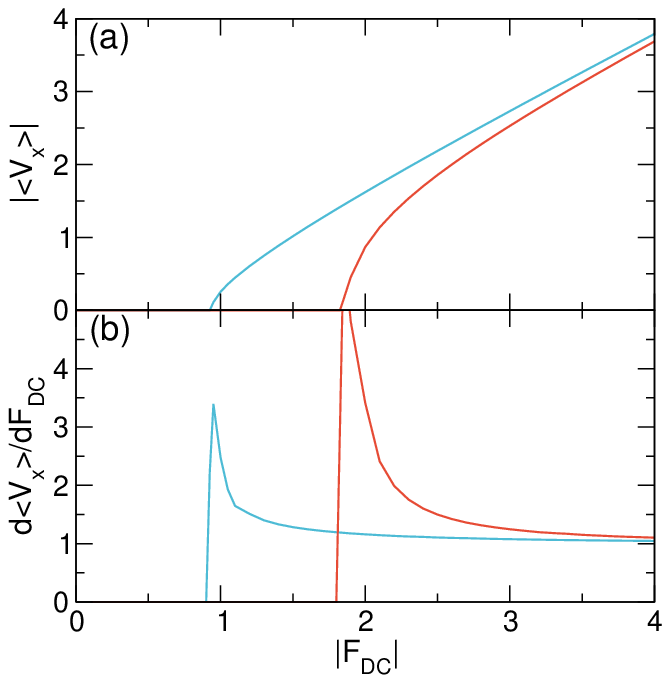}
\caption{Diode effect in a system under dc driving
with $N_e=10$ charges coupled to the substrate
with $a=2.117$ illustrated
in Fig.~\ref{fig:1}.
Here
$A_p = 1.25$, $F^{\rm hard}_{p} = 1.875$,
$F^{\rm easy}_{p} = 0.94$,
and $\rho=0.007$,
so the response is close to the single-particle limit.
(a) $|\langle V_{x}\rangle|$ vs $|F_{DC}|$ for driving in the easy
direction, $+x$ (blue), and hard direction, $-x$ (red).
(b) The corresponding $d\langle V_{x}\rangle/dF_{DC}$
vs $|F_{DC}|$ curves showing a finite depinning threshold
and a nonlinear sliding regime.
}
\label{fig:2}
\end{figure}

We primarily focus on the non-thermal case
and first examine the collective effects
on diode behavior under dc driving in the $+x$ or $-x$ direction.
In Fig.~\ref{fig:2}(a,b) we plot $|\langle V_{x}\rangle|$ versus
$|F_{DC}|$ and $dV_{x}/dF_{DC}$ for a system with a
substrate lattice spacing
of $a = 2.117$ for only $N_e=10$ particles, producing behavior
close to the single particle limit.
Here $A_p = 1.25$, so the maximum pinning force for driving in the hard
direction is $F^{\rm hard}_{p} = 1.875$
and that for driving in the easy direction
is $F^{\rm easy}_{p} = 0.94$.
The system shows a simple diode effect
in which the depinning threshold
falls at the maximum pinning force for the direction of drive being used.
Thus, the depinning threshold is larger for hard direction driving.
Within the sliding phase, for a given value of $|F_{DC}|$
the magnitude of the velocity is higher for driving in the easy direction than
for driving in the hard direction,
but this velocity difference decreases with increasing
$|F_{D}|$ and is significant
only in the nonlinear sliding regime.
The velocity-force curves increase linearly
with increasing $|F_D|$ at high drives, as
shown in Fig.~\ref{fig:2}(b).
In principle, if an ac drive is applied to this system,
a ratchet effect should
occur whenever the ac driving amplitude is larger than $F_p^{\rm easy}$ but
smaller than $F_p^{\rm hard}$;
however, since the velocities for the two driving directions
are also different in the sliding regime,
a ratchet effect should
continue to occur as long as the ac drive amplitude falls above
$F_p^{\rm easy}$ but within the nonlinear velocity-force regime.
We can compare our dc transport results to the work of
Zakharov {\it et al.} \cite{Zakharov19}, who
also found that there is a diode effect in the depinning thresholds
and that there is a region above the higher depinning threshold 
where the magnitude of the velocity is lower for driving in the hard
direction than for driving in the easy direction.
Zakharov {\it et al.} also observed
that at high drives, the
magnitude of the velocities in the sliding regime
for driving in the two directions gradually approach the same value
with increasing drive.
At high electron densities
when there are multiple electrons per well,
Zakharov {\it et al.} found evidence for a two-step depinning transition;
however, they
did not measure the number of defects in the system.

\begin{figure}
\includegraphics[width=\columnwidth]{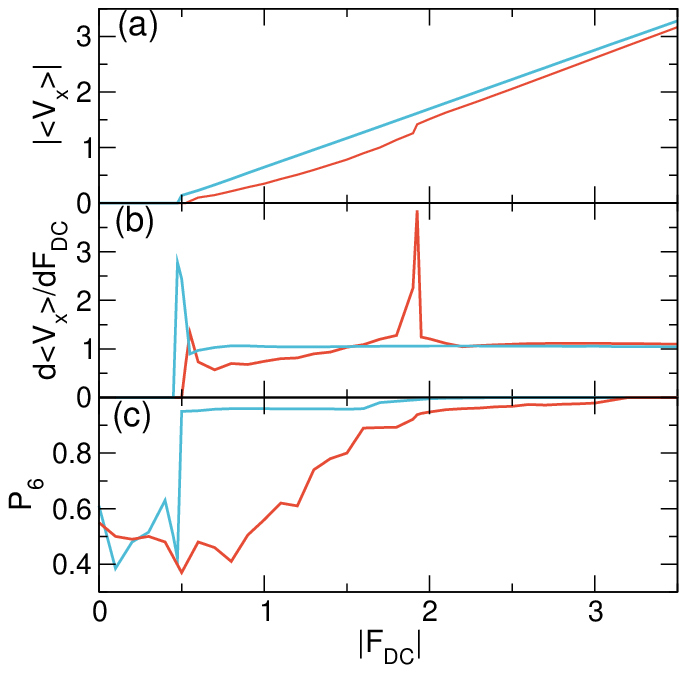}
\caption{ 
Diode effect for the same system from Fig.~\ref{fig:2}
with dc driving, $a=2.117$, and $A_p=1.25$
but with $N_{e} =  864$, giving $\rho=0.67$.
(a) $|\langle V_{x}\rangle|$ vs $|F_{DC}|$
for driving in the easy direction, $+x$ (blue),
and the hard direction, $-x$ (red),
showing a diode effect.
(b) The corresponding $d\langle V_{x}\rangle/dF_{DC}$ vs
$|F_{DC}|$ curves showing the
finite depinning threshold and the nonlinear sliding regime.
There is a two-step
depinning threshold for driving in the hard direction.
(c) The corresponding fraction of sixfold-coordinated particles
$P_{6}$ vs $|F_{DC}|$,
showing an extended region of disordered
flow for driving in the hard direction.
}
\label{fig:3}
\end{figure}

\begin{figure}
\includegraphics[width=\columnwidth]{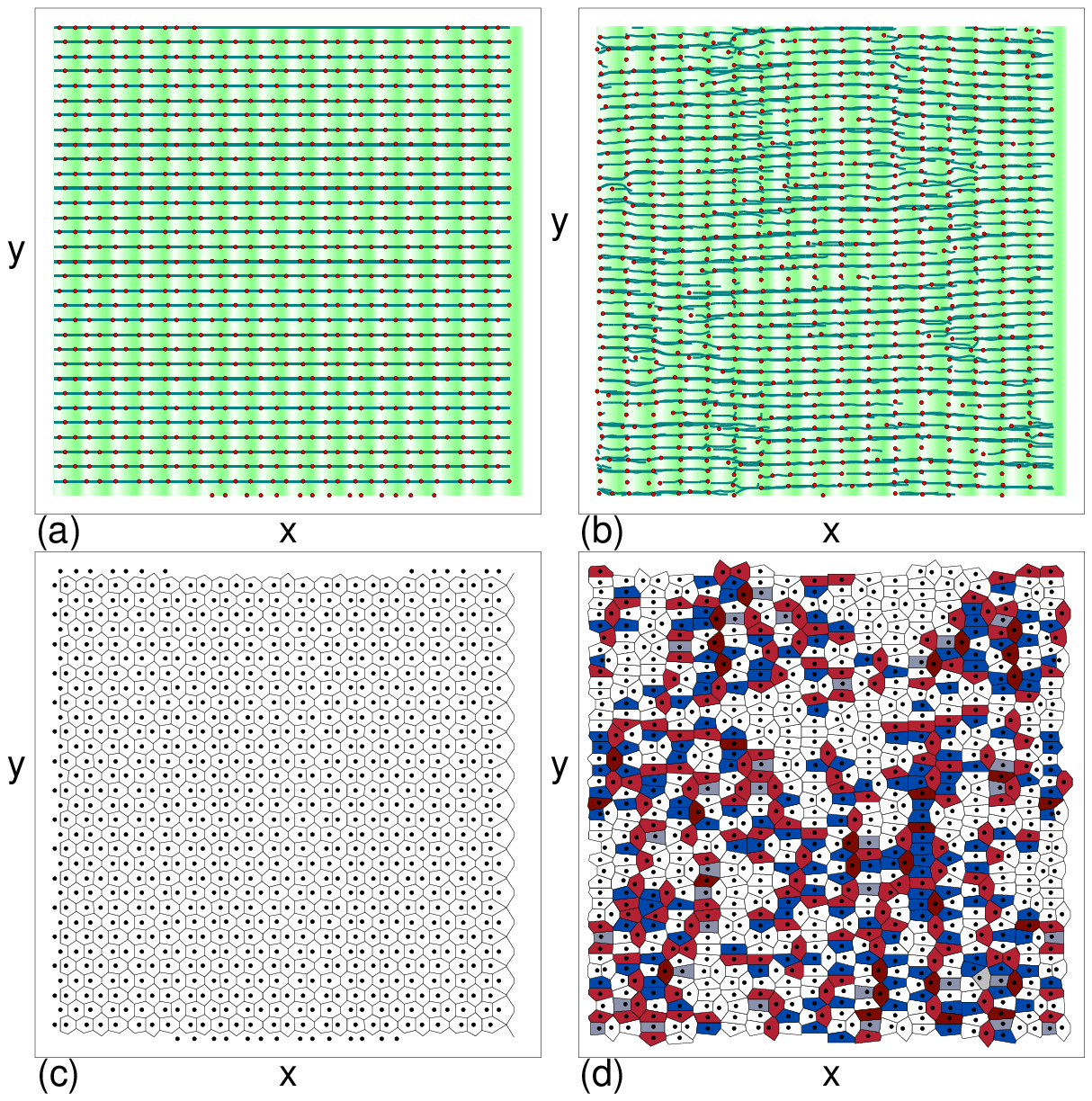}
\caption{
(a,b) Images of the asymmetric substrate (green),
the localized electrons (red dots), and the electron trajectories over
a fixed period of time (blue lines)  
for the system from Fig.~\ref{fig:3} with dc
driving, $a=2.117$, $A_p=1.25$, and $N_e=864$.
(a) Driving in the easy direction with $F_{DC} = 0.9$. The
flow is ordered.
(b) Driving in the hard direction
with $F_{DC} = -0.9$. The flow is disordered.
(c) The Voronoi construction for the particle positions in panel (a) shows
a weakly modulated
triangular lattice.
(d) The Voronoi construction for the particle positions in panel (b)
shows strong disorder.
In panels (c) and (d), polygon colorings indicate particle coordination
numbers of six (white), five (blue), seven (light red), four (gray), and
eight (dark red).
}
\label{fig:4}
\end{figure}

We next consider a denser system with an electron lattice spacing of
$a_{e} = 0.816$ and $N_e=864$ charges, so that $a_{e}/a  = 1.73$.
In this case, there are multiple electrons per substrate minimum.
In Fig.~\ref{fig:3}(a), we plot
$|\langle V_{x}\rangle|$ versus $|F_{DC}|$
for driving in the easy and hard directions.
Figure~\ref{fig:3}(b) shows the corresponding
$d\langle V_x\rangle/dF_{DC}$ versus $|F_{DC}|$
curves, while in Fig.~\ref{fig:3}(c)  we plot
the fraction of particles with six neighbors, $P_{6}$,
versus $|F_{DC}|$.
We obtain $P_6$ from a Voronoi construction, $P_6=N_e^{-1}\sum \delta(z_i-6)$,
where $z_i$ is the coordination number of particle $i$.
$P_{6} = 1.0$ indicates a triangular lattice.
For $\rho=0.67$ in Fig.~\ref{fig:3},
the depinning thresholds are $F_c^{\rm easy}=0.45$ for driving in the 
easy direction and
$F_c^{\rm hard}=0.525$ for driving in the hard direction,
indicating that the collective electron-electron interactions have
strongly reduced the depinning threshold
for both hard and easy driving well below the
maximum force exerted by the substrate in each case.
This occurs because there are multiple charges sitting in each
substrate trough,
so any given charge experiences
both the driving force and repulsive forces from the
neighboring charges.
Even though the depinning thresholds are
diminished by the increase in electron-electron interactions,
there are still extended regions
of drive amplitude for which
$|\langle V_{x}\rangle|$ is considerably lower for driving in the
hard direction than for driving in the easy direction. 
There are transitions as a function of drive in the collectively
interacting system for driving in the easy direction.
The first transition occurs from a pinned state 
containing topological defects, as indicated by the low
values of $P_{6}$ at low drives,
to an ordered sliding triangular lattice.
This transition produces
a single peak in the 
$d\langle V_{x}\rangle/dF_{DC}$ curves and
is accompanied by an increase in $P_{6}$
to a value $P_6 \approx 1$ just above depinning.
In Fig.~\ref{fig:4}(a), we illustrate the particle 
locations and trajectories for the system
from Fig.~\ref{fig:3} under an easy direction
dc drive of $F_{DC} = 0.9$.
The particles form a triangular lattice that
moves as an elastic solid with
ordered and straight trajectories.
Figure~\ref{fig:4}(c) shows the corresponding Voronoi construction,
where all of the particles have six neighbors and
a triangular solid forms with a weak density modulation.
For driving in the hard direction with $0.5 < |F_{DC}| < 1.95$,
the velocity is considerably lower 
than for driving in the easy direction,
and the velocity-force curves are nonlinear.
Over this range of $|F_{DC}|$, the flow is disordered,
as illustrated in Fig.~\ref{fig:4}(c), and
a portion of the particles are temporarily pinned while the remaining
particles flow. This results in the emergence of
a number of topological defects, as shown in the
corresponding Voronoi construction in Fig.~\ref{fig:4}(d).  
For $|F_{DC}|>1.95$,
the system becomes dynamically ordered for driving in the hard direction,
producing a jump up in $|\langle V_{x}\rangle|$ in Fig.~\ref{fig:3}(a);
however, even above the dynamic reordering transition,
the velocity for driving in the hard direction remains lower
than for driving in the easy direction.
Dynamic ordering occurs when $|F_{DC}|$ is greater than
the maximum pinning force in the hard direction,
allowing all of the particles to flow at the same time.
Similar dynamical ordering at high drives has been observed
in numerous other driven systems that exhibit depinning for systems with
random disorder \cite{Koshelev94,Moon96,Ryu96,Olson98a,Pardo98,Reichhardt17}
as well as for systems with
1D \cite{Reichhardt05a,Reichhardt16a}
and 2D periodic substrates \cite{Reichhardt97,Gutierrez09,Reichhardt17}.

In general, dynamical ordering occurs in strongly interacting
particle assemblies
with pinning once the drive is high enough that
all the particles are flowing, allowing the system to act more elastically.
Under these conditions, the particle-particle interactions can induce the
formation of the ordering that the system would have had
in the absence of a substrate.
The nonlinear regime for driving in the hard direction
is accompanied by two peaks in the $d\langle V_{x}\rangle/dF_{D}$ curves
in Fig.~\ref{fig:3}(b), where the first peak is associated with
depinning and the second with the dynamic ordering.
Only one peak occurs for driving in the easy direction, since for this
driving direction the depinning and ordering occur simultaneously.
The results in Figs.~\ref{fig:3} and \ref{fig:4} suggest that a
ratchet effect should appear in the strongly interacting
Wigner crystal regime,
while a diode effect should be possible for
a Wigner crystal coupled to asymmetric substrates for a range of fillings.
For the case of dc driving, we only consider a few fillings;
however, in general, we expect that the depinning thresholds and
magnitude of the diode effect will show oscillations due to
commensurability effects,
similar to what has been seen for
other two-dimensional Wigner crystal systems on one-dimensional
periodic substrates \cite{Moskovtsev20}.
Such commensurability effects have been observed previously
in a range of other two-dimensional assemblies of
interacting particles on 1D symmetric substrates,
such as vortices in type-II superconductors
\cite{Daldini74,Levitov91,Jaque02,Dobrovolskiy12,LeThien16},
magnetic bubbles \cite{Hu97},
and colloidal particles \cite{Zaidouny13}.
For example, Lu {\it et al.} \cite{Lu07}
found oscillations in the dc depinning threshold
for driving in both the easy and hard directions
for 2D systems of superconducting
vortices interacting with asymmetric 1D substrates.

\section{Ratchet Effects Under ac drives}

\begin{figure}
\includegraphics[width=\columnwidth]{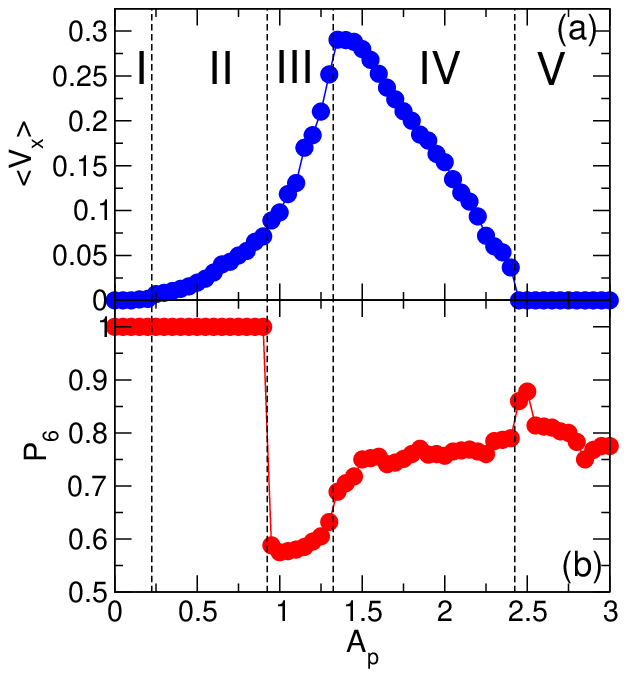}
\caption{The Wigner ratchet effect. 
(a) $\langle V_{x}\rangle$ vs substrate strength $A_{p}$ for  
a system with $N_e=672$ or $\rho=0.519$
under an
ac drive with $\omega = \omega_0$ and $F_{AC} = 1.8$.
The velocities are
averaged over fifty ac drive cycles.
The phases are: I, floating; II, elastic ratchet;
III, plastic ratchet; IV, diode ratchet; and V, pinned.
(b) The corresponding fraction of particles with six neighbors
$P_6$ vs $A_p$. 
}
\label{fig:5}
\end{figure}

We next examine the effect of applying an ac drive to
a Wigner crystal in the collective limit
with $N_{e} = 672$ and $a=2.117$,
which has the same dc behavior as the diode
system illustrated in Figs.~\ref{fig:3} and \ref{fig:4}.
We apply ac drive with frequency
$\omega = \omega_0$ and amplitude
$F_{AC} = 1.8$.
In Fig.~\ref{fig:5}(a) we plot 
$\langle V_{x}\rangle$ versus the substrate strength $A_{p}$,
where we average
the velocities over $50$ ac drive cycles.
Figure~\ref{fig:5}(b) shows the corresponding
$P_{6}$ versus $A_p$.
For $A_{p} < 0.925$ the system is in an elastic regime and the particles
form a triangular crystal for all portions of the ac drive cycle,
as indicated by the fact that
$P_{6} = 1.0$ for $A_{p} < 0.925$.
$\langle V_x\rangle$ is finite in this regime so there is still
a ratchet effect occurring.
We call this phase
II or the elastic ratchet state
since the system behaves like an elastic object coupled
to a ratchet substrate.
For this ac frequency and amplitude, the particles move over multiple
periods of the substrate during each half of the ac drive cycle.
When $A_{p} < 0.25$, the ratcheting motion
drops almost to zero
when the system undergoes an Aubry transition and enters
a floating phase \cite{Aubry78,Peyrard83,Brazda18,Shepelyansky19}, 
which we call phase I.
For $0.925 < A_{p} < 0.14$ the system
becomes disordered during both directions of the ac drive cycle,
as shown by the dip in $P_{6}$, and enters what we call 
a plastic ratchet or phase III, 
where the particles are moving over the barriers in
both directions but there is 
considerably more motion in the easy flow direction as indicated by the
growth of $\langle V_{x}\rangle$.
For $1.4 < A_{p} < 2.475$ the flow is only in the
positive direction and particles do not jump over the barriers
in the hard direction. We call this a diode ratchet or phase IV, and the
flow is still disordered in this regime. 
When $A_{p} > 2.475$, the particles remain pinned during both halves
of the ac cycle, and we term this
phase V or the pinned phase.
Boundaries between these phases produce signatures in
both the average velocity and $P_{6}$;
however, $P_{6}$ is a time-averaged quantity and does not
entirely capture the considerable structural changes that can occur
during a single ac drive cycle depending upon whether the response
is in phase II, III, or IV.

\begin{figure}
\includegraphics[width=\columnwidth]{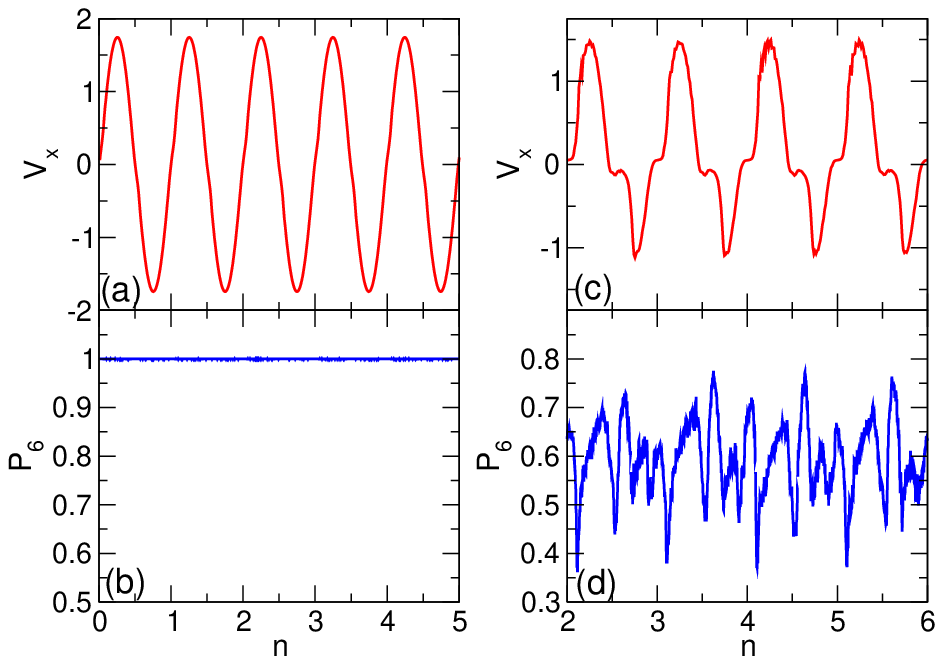}
\caption{
(a) $V_x$ vs time as a function of ac cycle number $n$
for the system in Fig.~\ref{fig:5} with $N_e=672$ or $\rho=0.519$
at $A_{p} = 0.65$ 
under an ac drive with
$\omega=\omega_0$ and $F_{AC}=1.8$, placing the system in
phase II or the elastic ratchet state.
(b) The corresponding $P_{6}$ vs $n$.
(c) $V_x$ vs $n$ for the same system at $F_{AC}=1.2$ in the
plastic ratchet phase III, where the particles are partially
disordered throughout the ac drive cycle.
(d) The corresponding $P_6$ vs $n$.
}
\label{fig:6}
\end{figure}

In Fig.~\ref{fig:6}(a) we plot the time series of $V_x$ versus
ac cycle number $n$
for the system in Fig.~\ref{fig:5} 
at $F_{AC} = 0.65$, while Fig.~\ref{fig:6}(b) shows
the corresponding $P_{6}$ versus $n$. Here the system
remains a crystal during all portions of the
ac drive cycle and the velocities have a nearly sinusoidal signature.
Since the lattice is more compressed when it is driven in the
hard direction, the lattice develops local density gradients on
the length scale of the substrate potential that tend to 
move the lattice in the easy flow direction to create
the phase II elastic ratchet.
The velocity time signature has a similar appearance in phase I (not shown)
except that the slight asymmetry that leads to the ratchet effect
disappears.
Figure~\ref{fig:6}(c) shows $V_{x}$ versus $n$ for 
$F_{AC} = 1.2$ where the system is in phase III or the plastic
ratchet state.
Here it is easy to see that
the velocities are larger during the $+x$ portion of the ac drive
cycle.
The system is disordered during both halves of the ac drive cycle
and only a portion of
the particles are able to hop in either direction;
however, as shown in the plot of $P_6$ versus $n$ in 
Fig.~\ref{fig:6}(d), the amount of topological order present
strongly depends on the stage of the ac drive cycle.
The system oscillates from a maximum of
$P_{6} = 0.76$ to a minimum of $P_{6} = 0.37$,
and the particles are the most disordered at the point where
the velocity is lowest or where the
greatest number of particles are pinned.  
Due to the asymmetry of the substrate,
the amount of six-fold ordering does not follow a simple sinusoidal
oscillation,
and there are peaks and valleys
in $P_6$ corresponding to when the
particles are slow or just starting to move again.
We find that $P_{6}$ does not follow exactly the
same path during each cycle due to the
plastic or disordered nature of the flow,
which permits
particles to exchange positions with each other and
generate a moderately chaotic flow. 

\begin{figure}
\includegraphics[width=\columnwidth]{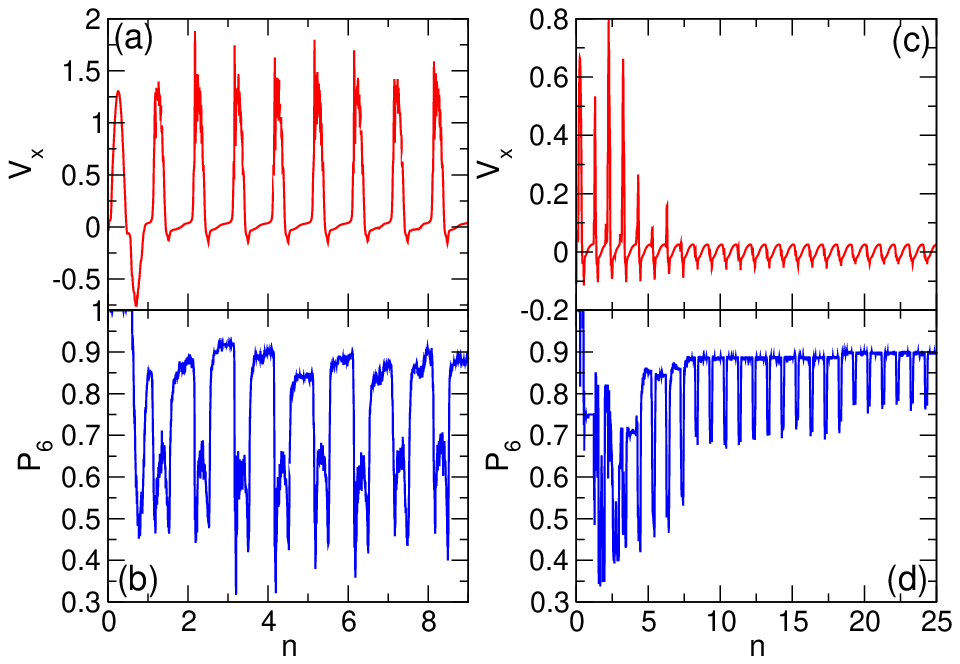}
\caption{
(a) $V_x$ and (b) $P_{6}$ vs $n$
for the system in Fig.~\ref{fig:5}
with $N_e=672$ or $\rho=0.519$
at $A_{p} = 1.5$ under an ac drive with $\omega=\omega_0$ and
$F_{AC}=1.8$
in the diode ratchet phase IV,
where the motion is only in the easy flow direction.
(c) $V_x$ and (d) $P_6$ vs $n$
for the same system in the pinned phase V at $A_{p} = 2.5$, 
where there is an initial transient before the particles settle
into the pinned state.
}
\label{fig:7}
\end{figure}

\begin{figure}
\includegraphics[width=\columnwidth]{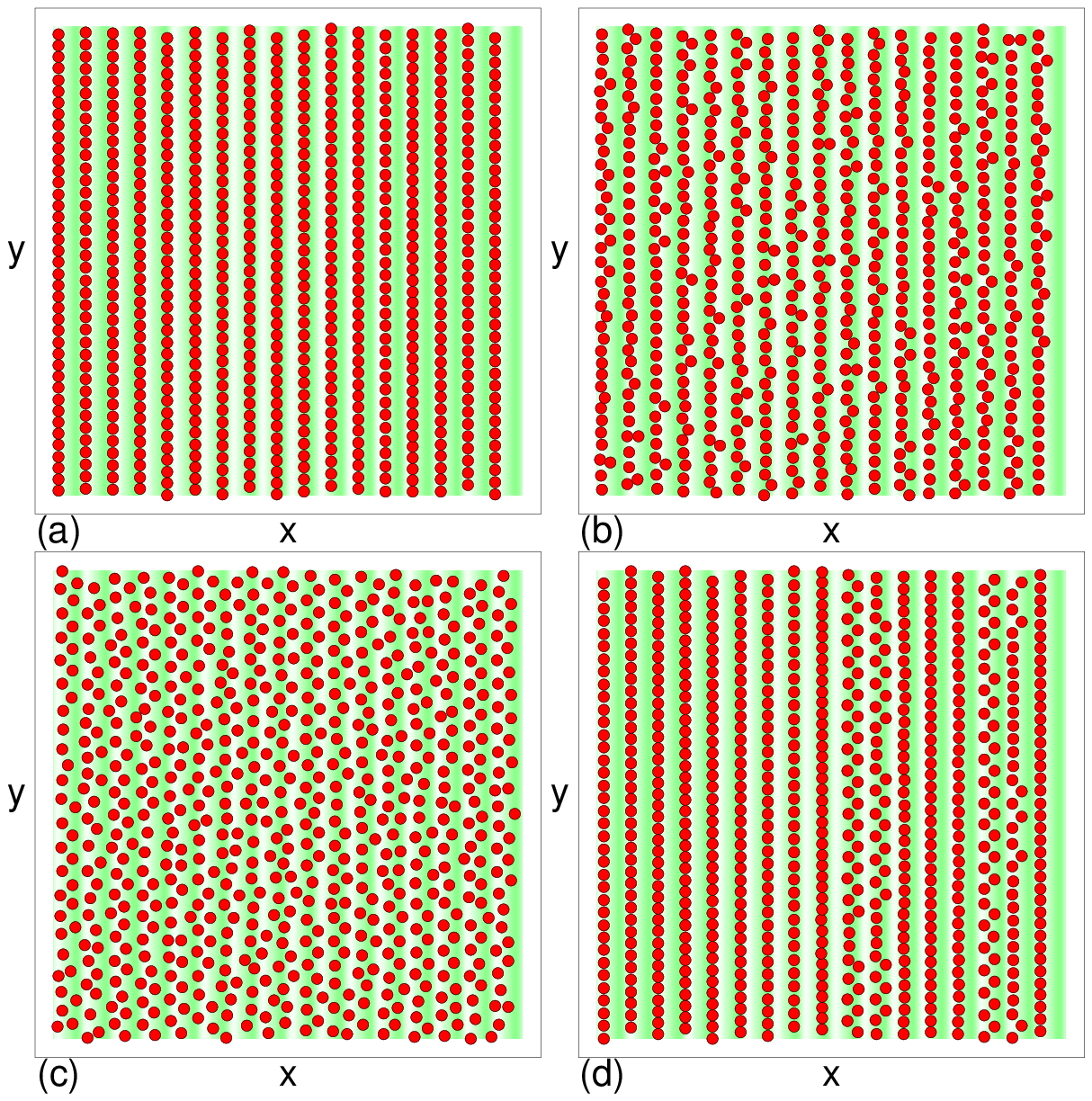}
\caption{Images of the asymmetric substrate (green) and the particle
positions (red) for the system from Fig.~\ref{fig:5} with
$N_e=672$ or $\rho=0.519$
under an ac drive with
$\omega=\omega_0$ and $F_{AC}=1.8$.
(a,b,c) The diode ratchet phase IV at $A_{p}=1.5$.
(a) The part of the ac drive cycle where $P_{6}$ is high, showing
the formation of 1D structures.
(b) The part of the ac drive cycle where $P_{6}$ is the lowest,
showing that the particles are starting to hop in 
the easy flow direction.
(c) The part of the ac drive cycle where the driving is highest in
the $+x$ direction
and $P_{6}$ has an intermediate value.
(d) The pinned configurations for phase V at $A_{p}=2.6$
during the portion of the ac drive cycle where some of the particles form 
a zig-zag pattern instead of a 1D structure.
Throughout the full ac drive cycle,
the system jumps between pinned
purely 1D chains similar to those shown in panel
(a) and the pinned zig-zag state shown in panel (d).
} 
\label{fig:8}
\end{figure}

In Fig.~\ref{fig:7}(a) we plot $V_x$ versus $n$
for the system from Fig.~\ref{fig:5}
at $F_{AC} = 1.5$ in phase IV or the ratchet
diode phase, while Fig.~\ref{fig:7}(b) shows
the corresponding $P_{6}$ versus $n$.
There is transient motion in the $-x$ direction during
the first ac drive cycle, but then the system settles into a state
where the particles only move in the $+x$ or easy flow
direction.
Here $P_{6}$ is high when
the drive is in the hard flow
direction and lowest just at the point when
the particles begin to hop over the barriers. 
During the entire ac drive cycle,
$P_6$ oscillates from $P_6 \approx 0.49$ to $P_6=0.92$,
and exhibits an intermediate value near $P_6=0.64$. 

In Fig.~\ref{fig:8}(a) we illustrate the particle positions
for the system in Fig.~\ref{fig:7}(a) during
the $-x$ portion of the ac drive
cycle where the particles form 1D chains 
and $P_{6}$ is high since the particles
have six neighbors on average. 
The chain formation results when the particles accumulate
against the hard direction side of the potential. 
Figure~\ref{fig:8}(b) shows the particle configuration
when $P_{6}$ reaches its lowest value during the ac cycle,
which correlates with the moment when
particles start hopping over the barriers in the $+x$ direction.
In this case, the flow is plastic since some particles are moving and
others are not, resulting in neighbor exchange and
leading to the generation of topological defects.
In Fig.~\ref{fig:8}(c) we show the particle positions during the
$+x$ portion
of the ac drive cycle
when most of the particles are moving in the easy flow
direction.
According to Fig.~\ref{fig:5},
$A_{p} = 1.5$ is close to the substrate strength value for which
particles begin to move only in the $+x$ direction and
the maximum ratchet effect occurs.
As $A_{p}$ is further increased,
the flow is still only in the $+x$ direction but
not all of the particles are able to jump out of their substrate minima
during each ac drive cycle,
reducing the ratchet efficiency.

In Fig.~\ref{fig:7}(c,d) we plot $V_{x}$ and 
$P_{6}$ versus cycle number $n$
for a system with $A_{p} = 2.5$.
Here, some initial motion resembling phase IV flow occurs,
but after seven ac drive cycles,
the system settles into a
phase V or pinned state where the average 
$V_{x} = 0.$
Even though there is no net flow in this phase,
there are still some oscillations in $P_{6}$.
This is due to the pinned configuration 
jumping between purely 1D
chains and the zig-zag pattern illustrated in
Fig.~\ref{fig:8}(d).
As $A_{p}$ is further increased,
the lifetime of the transient flow decreases
and the particle configurations become more 1D during
all portions of the ac drive cycle. 

\subsection{Varied Fillings}

\begin{figure}
\includegraphics[width=\columnwidth]{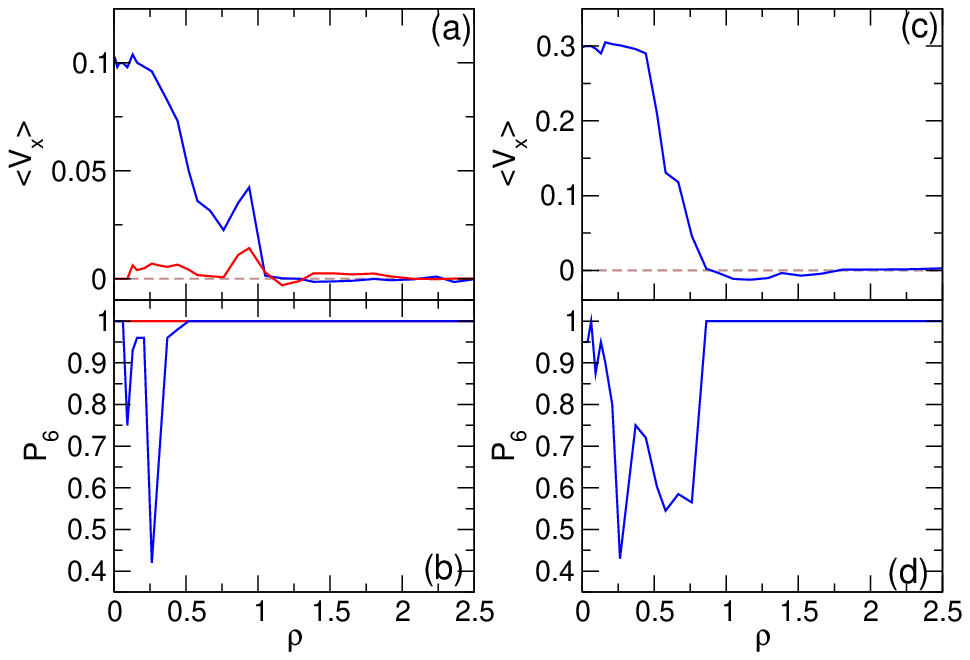}
\caption{
(a) $\langle V_{x}\rangle$ vs particle density $\rho$
for the system from Figs.~\ref{fig:6} and \ref{fig:7}
under an ac drive with $\omega=\omega_0$ and $F_{AC}=1.8$ at
$A_{p} = 0.75$ (blue) and $A_{p} = 0.25$ (red).
(b) The corresponding $P_{6}$ vs $\rho$.
(c) $\langle V_{x}\rangle$ and (d) $P_{6}$ vs $\rho$
for the $A_p=1.25$ system, where
there is a weak ratchet reversal at higher filings.
}
\label{fig:9}
\end{figure}

We next examine the evolution of the different ratchet effects for
varied fillings and substrate strengths
for the system in Figs.~\ref{fig:6} and \ref{fig:7}.
In Fig.~\ref{fig:9}(a), we plot
$\langle V_{x}\rangle$ versus particle density $\rho$
for $A_{p} = 0.75$ and $A_{p} = 0.25$, while
Fig.~\ref{fig:9}(b) shows the corresponding 
$P_{6}$ versus $\rho$ curves.
For $A_{p} = 0.25$, there is only a weak ratchet effect
with a local peak near $\rho = 1.0$, the density at which
the particles form a commensurate lattice that can more effectively
couple to the substrate.
At this substrate strength, a floating solid appears at other values
of $\rho$.
For $A_{p} = 0.75$, a triangular lattice is present when
$\rho \geq 0.5$. There is elastic ratchet motion
for $0.5 \leq \rho < 1.1$, while for higher $\rho$, the system 
is in a floating phase and the ratchet effect is close to zero. 
For $\rho < 0.5$, the system becomes partially plastic, and the
ratchet effect increases until it reaches a maximum efficiency
with $\langle V_x\rangle=0.1$ in the
low density single particle limit.
There is a peak in $\langle V_x\rangle$
near $\rho = 1.0$ due to formation of the commensurate lattice.
Fig.~\ref{fig:9}(c,d) shows $\langle V_x\rangle$ and $P_6$
versus $\rho$ for $A_{p} = 1.25$.
Here we find a plastic ratchet regime for $0 < \phi < 0.9$ where there is
a pronounced ratchet effect and the system is disordered.
For $0.9 \leq \rho < 2.0$
the system forms a triangular lattice
and there is a weak ratchet reversal as
indicated by the negative value of $\langle V_{x}\rangle$.
We note that 
ratchet reversals are commonly
observed in systems with collective interactions.
For $\rho \geq 2.0$ the system is in a floating phase
and the ratchet effect is almost zero.

\begin{figure}
\includegraphics[width=\columnwidth]{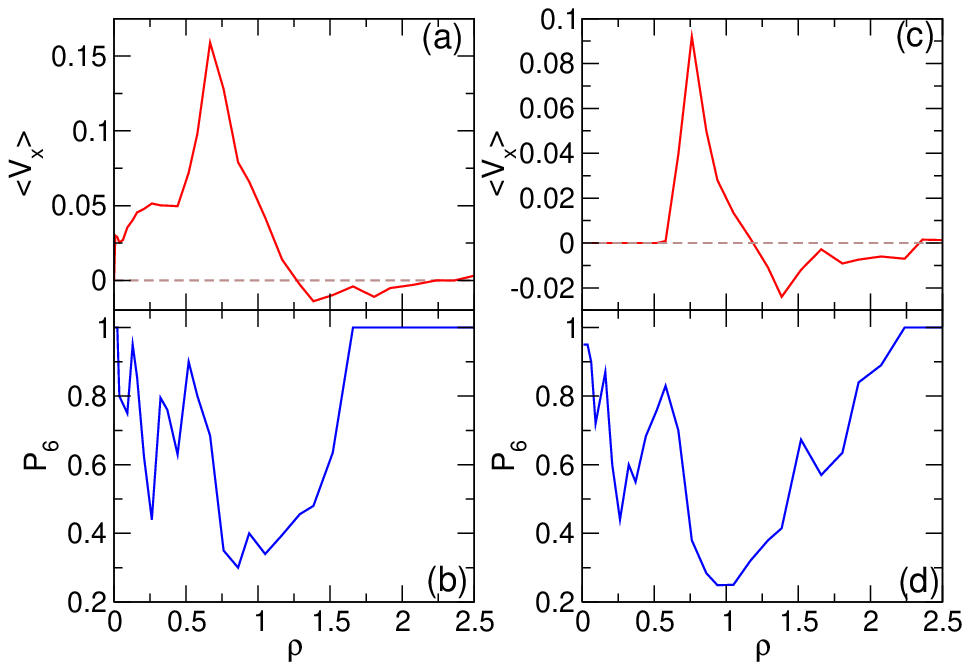}
\caption{(a) $\langle V_x\rangle$ and (b) $P_{6}$
vs $\rho$ for the system from Fig.~\ref{fig:9}
under an ac drive with $\omega=\omega_0$ and $F_{AC}=1.8$ at
$A_{p} = 2.25$. 
(c) $\langle V_{x}\rangle$ and (d) $P_{6}$
vs $\rho$ for the same system at $A_{p} = 2.75$. Here
there is a pinned phase for lower densities $\rho$. 
}
\label{fig:10}
\end{figure}

In Fig.~\ref{fig:10}(a,b) we
plot $\langle V_{x}\rangle$ and $P_{6}$
versus $\rho$ for the same system at $A_{p} = 0.75$.
The system is in the diode ratchet regime for $0.0 < \rho < 1.0$
and enters a
plastic ratchet regime for $1.0 \leq \rho < 1.5$.
The ratcheting motion in the plastic regime
can be in either the $+x$ or $-x$ direction.
The particles form a triangular solid for $\rho > 1.75$,
and the ratchet effect is close to zero for
$\rho > 2.25$.
Figure~\ref{fig:10}(c,d) shows $\langle V_x\rangle$ and $P_6$
versus $\rho$ at $A_{p} = 2.75$
where now there is a pinned phase for $\rho < 0.5$,
a diode ratchet, and a plastic ratchet.
The negative ratchet motion is clearly visible, but for
$\rho > 2.25$, the ratchet effect is lost.
Within the pinned phase, the amount of topological disorder decreases
with increasing $\rho$, and
$P_6$ passes through a local dip at the
crossover from the pinned state to the diode ratchet motion.

\begin{figure}
\includegraphics[width=\columnwidth]{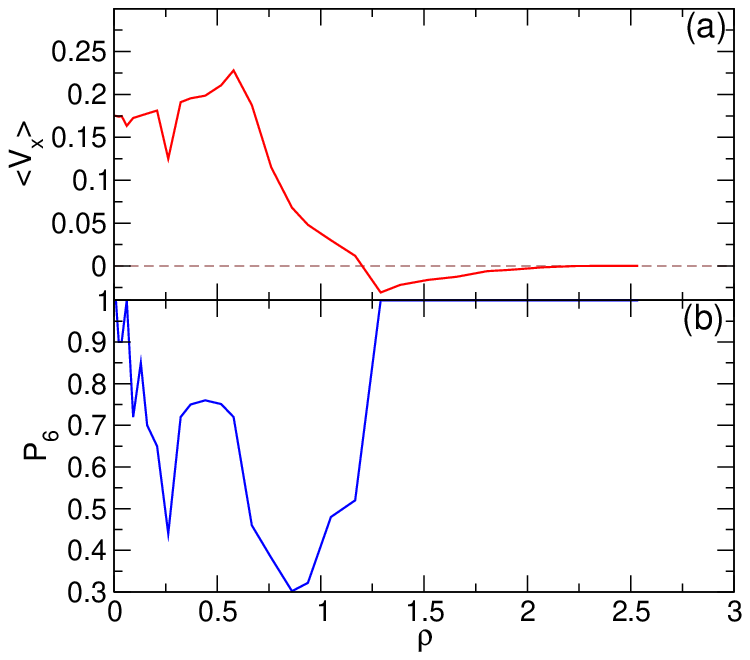}
\caption{
(a) $\langle V_{x}\rangle$ and (b) $P_{6}$ vs $\rho$
for the system in Fig.~\ref{fig:9} under an ac drive
with $\omega=\omega_0$ and $F_{AC}=1.8$ at  
$A_{p} = 1.75$. The transition to the elastic ratchet is associated
with an onset of reversed ratchet motion. Near $\rho = 0.265$ 
there is a dip in $P_{6}$ that corresponds to the formation of a
square commensurate lattice.
}
\label{fig:11}
\end{figure}

\begin{figure}
\includegraphics[width=\columnwidth]{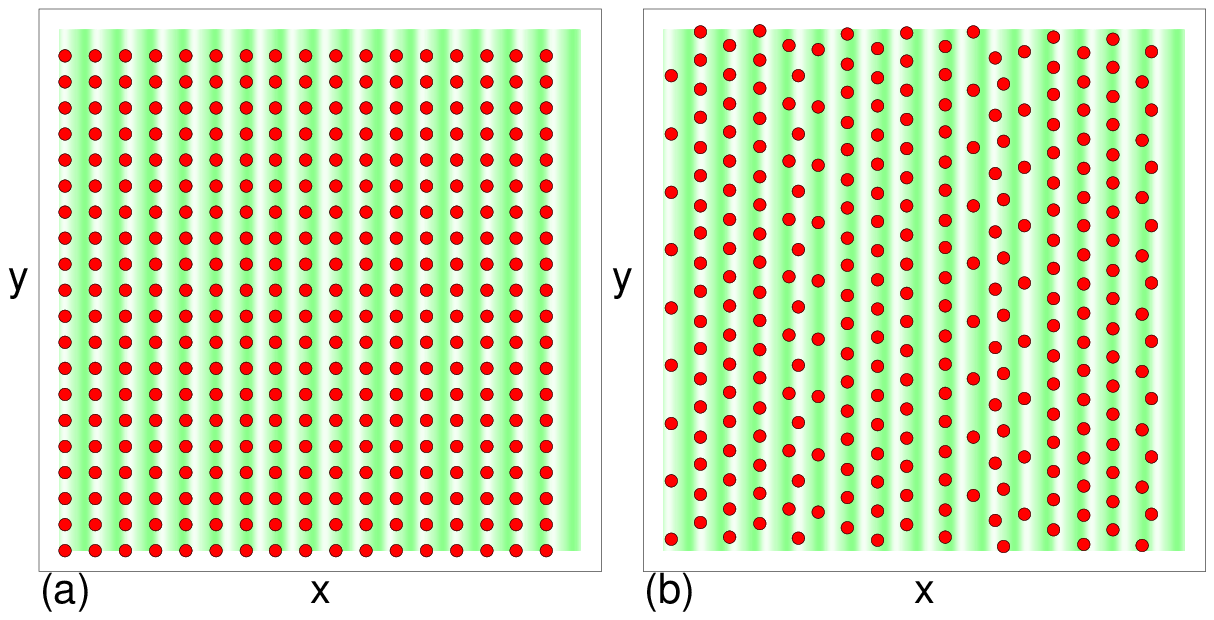}
\caption{Images of the asymmetric substrate (green) and the
particle positions (red) for the system from Fig.~\ref{fig:11}
under an ac drive with $\omega=\omega_0$ and $F_{AC}=1.8$ at $A_p=1.75$.
(a) $\rho = 0.265$ where the system forms a square lattice. 
(b) $\rho = 0.2$ where the system forms disordered triangular lattice. 
}
\label{fig:12}
\end{figure}

\begin{figure}
\includegraphics[width=\columnwidth]{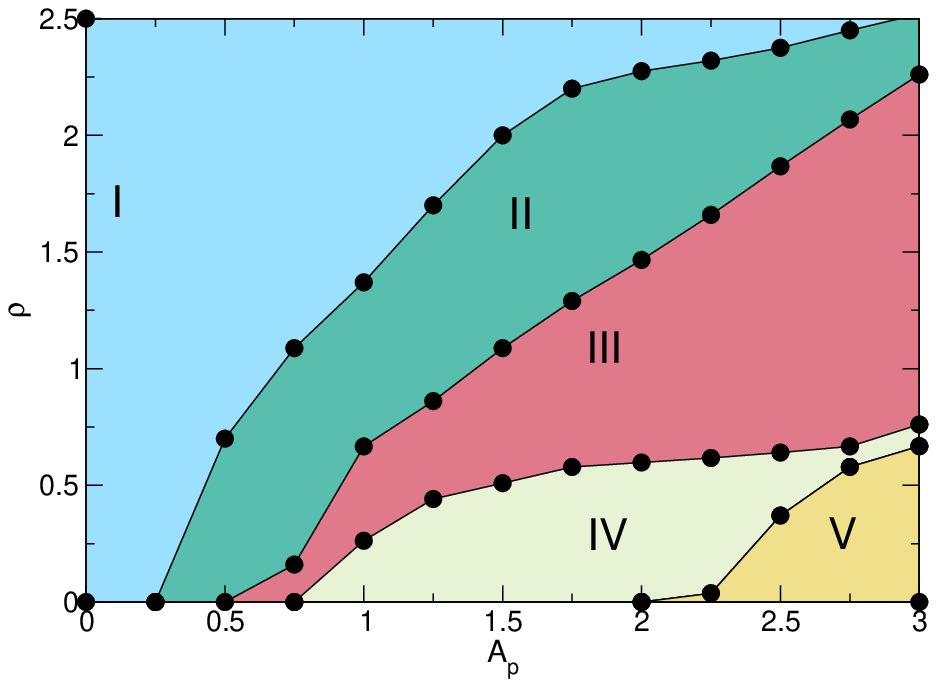}
\caption{
Dynamic phase diagram as a function of $\rho$ vs $A_p$ for the  
systems in Figs.~\ref{fig:9} through \ref{fig:12} under
ac driving with $\omega=\omega_0$ and $F_{AC}=1.8$ highlighting
the elastic or floating solid phase I,
the elastic ratchet phase II, the
plastic ratchet phase III where the system is disordered,
the diode ratchet phase IV, and
the pinned phase V.
}
\label{fig:13}
\end{figure}

In Fig.~\ref{fig:11}(a,b), we plot
$\langle V_{x}\rangle$ and $P_{6}$ versus $\rho$
for the same system as in Fig.~\ref{fig:9} but with  $A_{p} = 1.75$. 
We observe a number of distinct features.
The system is in the diode ratchet phase IV
for $0 < \rho < 0.55$,
and the maximum ratchet effect occurs near the crossover
from phase IV to phase III.
The onset of the plastic ratchet
phase III is associated
with a drop in $P_{6}$, and the particles undergo
disordered flow for both the negative and positive portions of the
ac drive cycle.
Near $\rho = 1.25$, there is a crossover 
from phase III to phase II that coincides with
a change to a negative ratchet effect.
Within phase IV, near $\rho=0.265$ there is a window of density
where the particles form a commensurate structure with square ordering.
The ratcheting is in the $+x$ direction throughout this window, and
the system remains ordered during the entire ac driving cycle with
no particles exchanging neighbors.
In contrast, neighbor exchange does occur during the $+x$ portion of
the ac drive cycle at other fillings within the diode ratchet regime.
In Fig.~\ref{fig:12}(a), we show the configuration of the 
particles at $\rho = 0.265$, where the system forms a square lattice.
We note that $P_{6}$ is not zero for the square lattice because the
Voronoi algorithm sometimes adds extremely short sides that cause
a fourfold polygon to become sixfold.
Figure~\ref{fig:12}(b) shows the configuration at $\rho = 0.2$,
where a disordered triangular lattice is present.
The formation of a square or distorted lattice for a 2D system on a
1D substrate has been observed
previously for colloidal systems \cite{Hu97,Zaidouny13}.
In general, particles that would naturally form a triangular lattice
with lattice constant $a_0$ in the
absence of the substrate can still form triangular lattices
even when coupled to 
a strong substrate for densities at which 
$a_0$ matches the distance $a$ between adjacent substrate troughs.
In some cases, if the two length scales do not match perfectly,
the system instead forms a square or slightly
distorted triangular lattice \cite{Hu97,Zaidouny13}.
At nonmatching fillings on strong substrates, the system will form
smectic-like states, a mixture of smectic and zig-zag patterns,
or states containing localized topological defects.

Based on the features
in $V_{x}$ and $P_{6}$, in Fig.~\ref{fig:13} we create a
phase diagram as a function of $\rho$ versus $A_{p}$
for the system in Figs.~\ref{fig:9} through \ref{fig:12}.
For weak pinning or high $\rho$, the system is in
the floating solid phase I and there is little or no ratcheting.
At high densities, the charges are more strongly interacting and
the elastic constant of the particle lattice increases,
diminishing the effectiveness of the coupling to the substrate
In real Wigner crystals, the effect of increasing the electron
density is likely to be subtle since the denser system would,
at some point, become more liquid-like,
so for very high $\rho$, we would expect an additional fluid-like phase
to appear.
At small $\rho$ and high $A_p$,
there is a pinned phase V where the particles do not jump out of
the barriers.
Phase IV generally occurs for small $\rho$ and
intermediate $A_{p}$, and the maximum ratchet effects generally
fall along the line dividing phases III and IV.
The extent of the plastic ratchet regime increases with increasing
$\rho$ and $A_{p}$.
The locations of the different phases also depend on the ac drive
amplitude. For example, phase V shifts to higher substrate strengths
for larger ac drive amplitudes.
The substrate lattice spacing, ac drive frequency, and temperature also
affect the extent of the phases.
It is possible to define additional phases,
such as an elastic diode version of region IV,
where the flow is in only one direction but plastic
rearrangements do not occur.
There can also be variants of regions III or II where
the ratchet effect is in the $-x$ direction.
Such reversed ratchet phases generally disappear once
$\rho > 1.2$ for the parameters shown in Fig.~\ref{fig:13}.

\subsection{ac Amplitude and Ratchet Reversal}

\begin{figure}
\includegraphics[width=\columnwidth]{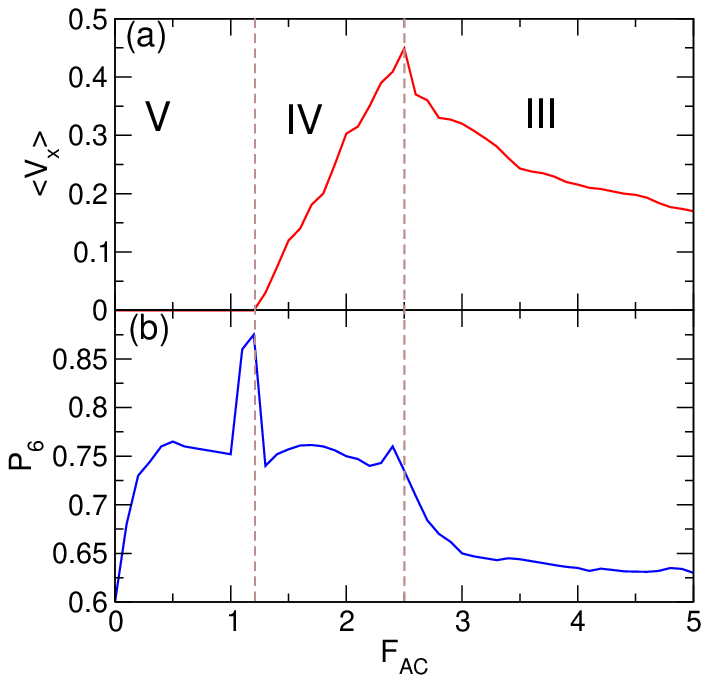}
\caption{
(a) $\langle V_{x}\rangle$ and (b) $P_{6}$ vs $F_{AC}$ for the system in
Fig.~\ref{fig:13}
at $\rho=0.515$ and $A_p=1.75$ under an ac drive with $\omega=\omega_0$.
}
\label{fig:14}
\end{figure}

In Fig.~\ref{fig:14}(a,b) we plot
$\langle V_{x}\rangle$ and $P_{6}$ versus
$F_{AC}$ for the system
in Fig.~\ref{fig:13} at $\rho = 0.515$ and
$A_{p} = 1.75$.
For $F_{AC}< 1.25$, the system
is in a pinned phase,
while for $1.25 < F_{AC} < 2.25$, the system is in phase IV.
At low values of $F_{AC}$, the pinned phase V has
smectic or stripelike ordering.
Near the crossover between phases V and IV,
the system becomes more topologically ordered, while in phase IV,
the system jumps
between a disordered flow state and a chain state.
The crossover from phase IV to phase III is marked by a peak
in $\langle V_x\rangle$ and falls slightly above a peak
in $P_6$.
For other values of $A_{p}$ we also find that there is an optimal
ac amplitude falling between phases IV and III where the ratchet effect
is maximized.

\begin{figure}
\includegraphics[width=\columnwidth]{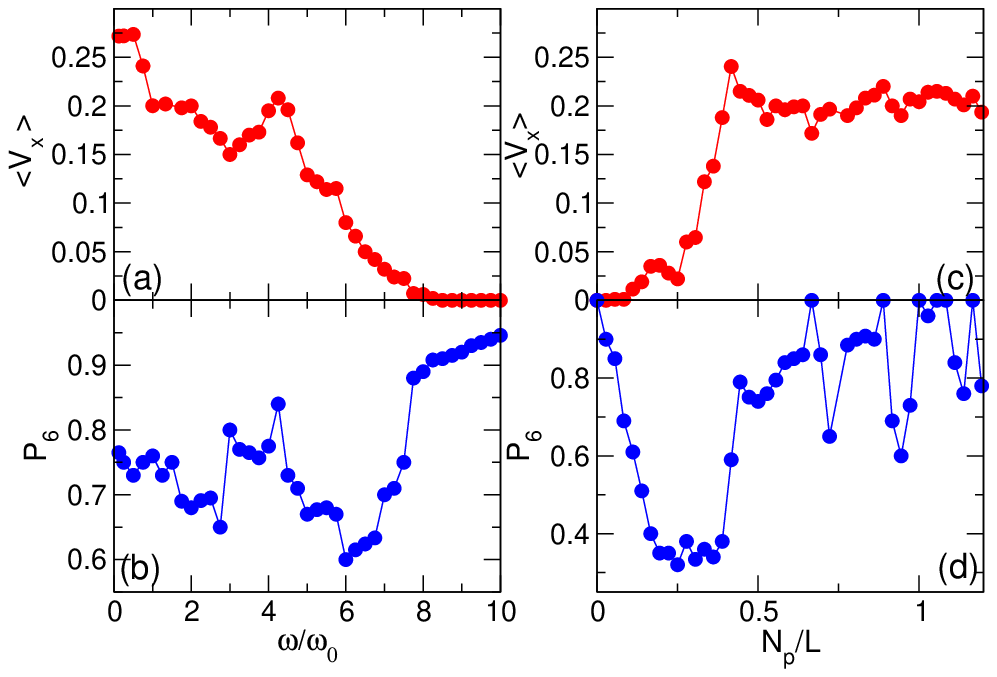}
\caption{
(a) $\langle V_{x}\rangle$ and (b) $P_{6}$ vs
$\omega/\omega_{0}$ in a system with
$F_{AC}  = 1.8$, $A_{p} = 1.7$, $N_p/L=0.472$, and
$\rho = 0.518$.
The ratchet efficiency generally decreases with increasing
$\omega$, but there
is a commensuration peak near $\omega/\omega_{0} = 4.5$.
(c) $\langle V_{x}\rangle$ and (d) $P_{6}$ vs $N_{p}/L$
in a system with $\omega/\omega_{0} = 1.0$,
$F_{AC}  = 1.8$, $\rho=0.518$, and $A_{p} = 1.7$.}
\label{fig:15}
\end{figure}

In Fig.~\ref{fig:15}(a,b), we consider a sample with
$A_p=1.7$, $N_p/L=0.472$, and $\rho=0.518$ where we fix the
ac drive amplitude to $F_{AC}  = 1.8$ and vary
$\omega/\omega_{0}$.
The ratchet efficiency generally
decreases with increasing $\omega/\omega_0$,
but passes through a peak near $\omega/\omega_{0} = 4.5$,
which also coincides with a peak in $P_{6}$.
For $\omega/\omega_{0} > 7.5$, the frequency is high
enough that the particles cannot respond
to the drive, the hopping rate over the barrier goes to zero, and
the system enters a 1D pinned configuration.
In Fig.~\ref{fig:15}(c,d), we fix $\omega/\omega_0=1.0$ for
the same sample and
change the periodicity or spacing $a=N_p/L$ of the underlying substrate
by varying the number $N_p$ of
pinning troughs.
High values of $N_p/L$ indicate a denser pinning array.
For $N_{p}/L < 0.45$, the system is in phase III, there 
are multiple rows of particles per trough,
and the ratchet efficiency drops,
while for $N_{p}/L < 0.2$, the particles cannot move far enough
during half of an ac drive cycle to jump over the
barriers and the system forms a triangular lattice between the troughs.
For $N_{p}/L > 0.45$, the ratchet efficiency increases when the
system enters phase IV.
The ratcheting can be plastic for part of the
ac drive cycle, leading to lower $P_{6}$ values;
alternatively,
when there is commensuration between the average particle
spacing and $N_p/L$, the ratcheting can be elastic and the particles
form a commensurate structure that moves in a diode-like
fashion without plastic deformation,
giving $P_{6} = 1.0$.
In general, when there are
many pinning troughs, there is a greater number of possible ways
to arrange the particles in order to
form a commensurate lattice.
These results show that more efficient ratchet
transport can occur when
the spacing of the substrate periodicity
is within a factor of two or less from the spacing of the
particles.

\begin{figure}
\includegraphics[width=\columnwidth]{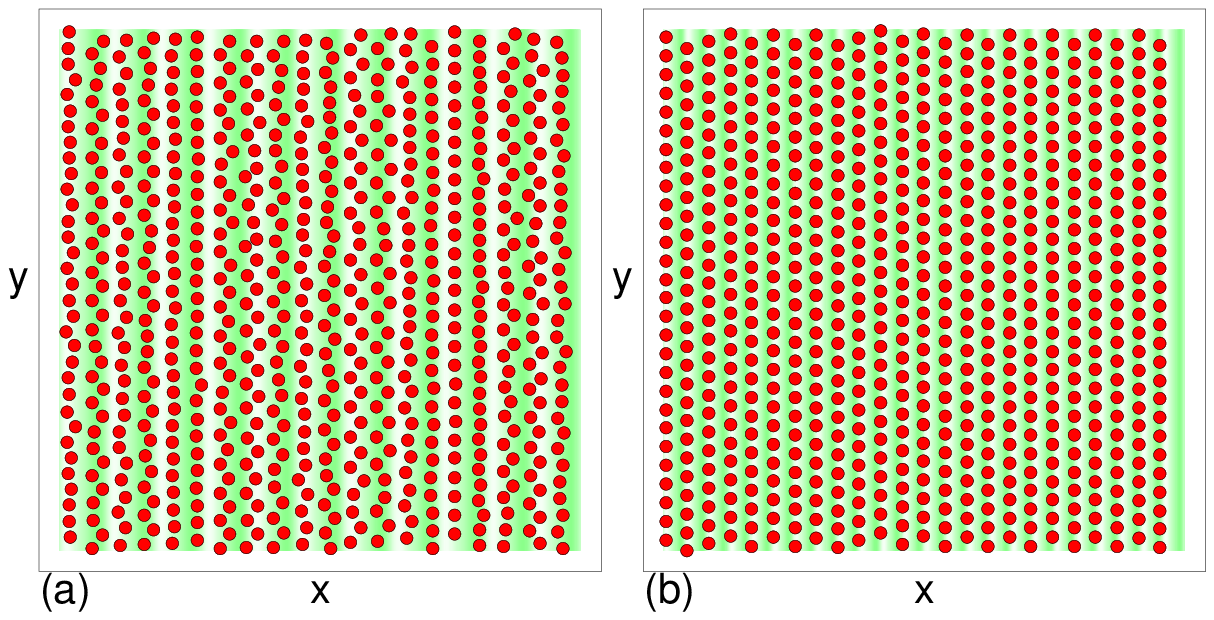}
\caption{Images of the asymmetric substrate (green) and the
particle positions (red) for the system in Fig.~\ref{fig:15}(c,d)
with $F_{AC}=1.8$, $A_p=1.7$, $\rho=0.518$, and $\omega/\omega_0=1.0$.  
(a) $N_{p}/L = 0.305$, where the system is
disordered.
(b) $N_{p}/L = 0.66$, where the system forms an ordered commensurate lattice.  
}
\label{fig:16}
\end{figure}

In Fig.~\ref{fig:16}(a) we show the configurations of the particles for
the system in Fig.~\ref{fig:15}(c,d) at
$N_{p}/L = 0.305$, where the
particles form a disordered structure and
the ratchet efficiency is low.
Figure~\ref{fig:16}(b) shows the
particle configurations for a commensurate case of
$N_{p}/L = 0.66$, where the system forms
a triangular lattice that undergoes
an elastic diode ratchet effect.

\begin{figure}
\includegraphics[width=\columnwidth]{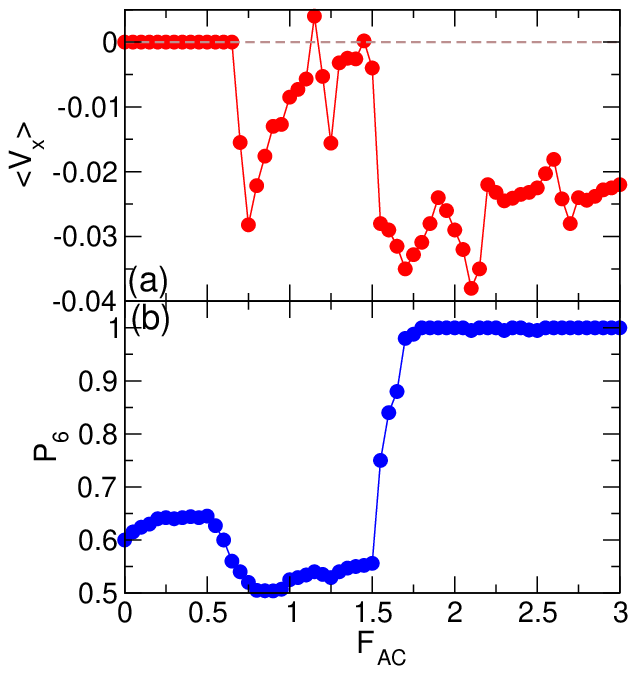}
\caption{(a) $\langle V_{x}\rangle$ and
(b) $P_{6}$ vs $F_{AC}$ for
a system with $\rho = 1.29$, $A_p = 1.75$, $\omega/\omega_{0} = 1.0$,
and $N_{p}/L = 0.472$.
A transition from a plastic to an elastic ratchet
occurs near
$F_{AC} = 1.5$.
}
\label{fig:17}
\end{figure}

We next examine the behavior of the negative ratchet, which occurs at high
particle densities.
In Fig.~\ref{fig:17} we plot $\langle V_{x}\rangle$ and $P_{6}$
versus $F_{AC}$ for
a system with $\rho = 1.29$, $A_p = 1.75$, $\omega/\omega_{0} = 1.0$,
and $N_{p}/L = 0.472$.
For $F_{AC} < 0.7$, the system is 
in a pinned state with multiple particles per pinning row.
For $0.75 < F_{AC} < 1.5$, the system is in a disordered or
plastic ratchet phase where the particles are
partially disordered for a portion of the ac drive cycle.
A small ratchet reversal appears near $F_{AC} = 1.1$.
For $F_{AC} > 1.5$, the system is in an elastic phase
and the reversed ratchet effect reaches its largest magnitude.

\begin{figure}
\includegraphics[width=\columnwidth]{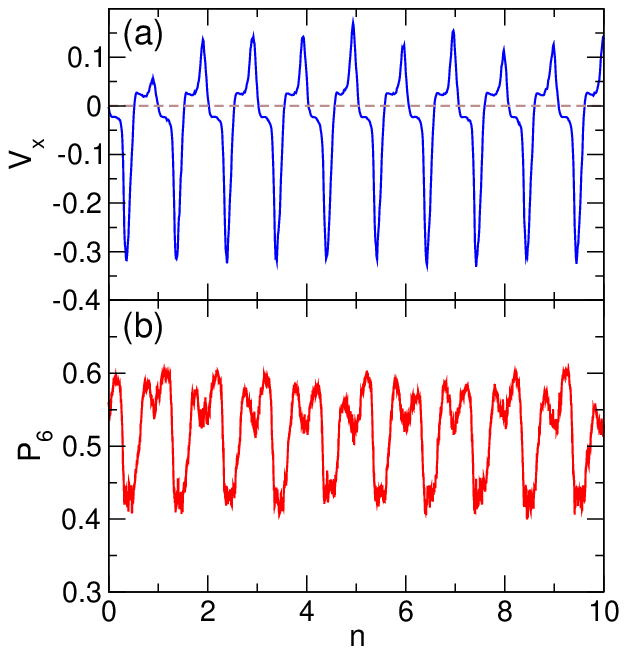}
\caption{ 
(a) $V_{x}$ vs cycle number $n$ and (b) $P_{6}$ vs $n$ for 
  the system from Fig.~\ref{fig:17}
  with $\rho=1.29$, $A_p=1.75$, $\omega/\omega_0=1.0$,
  and $N_p/L=0.472$
in the negative ratchet regime at $F_{AC} = 0.75$. 
}
\label{fig:18}
\end{figure}

In Fig.~\ref{fig:18} we plot
$V_{x}$ and $P_{6}$ versus the cycle number $n$ for the system from
Fig.~\ref{fig:17} in the negative ratchet regime at $F_{AC} = 0.75$.
The magnitude of the velocity is greatest during
the negative portion of the cycle, and the system is
at its most
disordered during the negative velocity peaks
when the particles undergo plastic flow.

\begin{figure}
\includegraphics[width=\columnwidth]{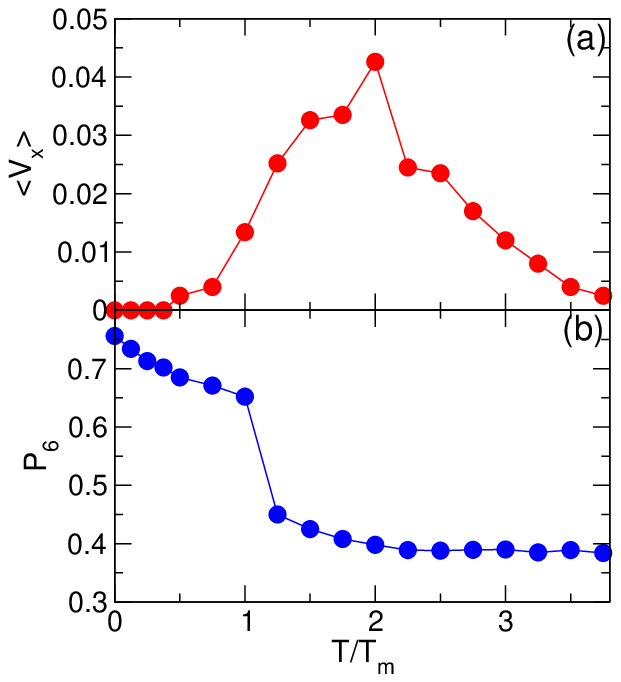}
\caption{ 
(a) $\langle V_{x}\rangle$ and (b) $P_{6}$ vs $T/T_{m}$ for a system
with $\rho = 0.518$, $A_{p} = 1.75$, $F_{AC} = 0.8$, and
$\omega/\omega_{0} = 1.0$.
$T_m$ is the temperature at which the substrate free system melts.
}       
\label{fig:19}
\end{figure}

We next consider the effects of thermal fluctuations on the ratchet
efficiency. Generally, if the substrate is strong, thermal
fluctuations can enhance the ratchet effect; however, the
ratchet efficiency is reduced when
the thermal fluctuations become large enough
to cause the particles to hop rapidly
in either direction on the substrate.
If the ac amplitude is already large enough that the
particles are moving for both directions of the ac drive
cycle, thermal fluctuations reduce the ratchet effect. 
In Fig.~\ref{fig:19} we plot
$\langle V_{x}\rangle$ and $P_{6}$ versus $T/T_{m}$ for a system
with $\rho = 0.518$, $A_{p} = 1.75$, $F_{AC} = 0.8$ and
$\omega/\omega_{0} = 1.0$.
For these parameters, at $T = 0.0$ the ac drive amplitude
is small enough that the system is in a pinned smectic phase.
We measure the temperature
in terms of the value $T_m$
at which a clean system with no substrate would melt.
Here the system is in a pinned phase or weak
ratchet regime up to $T/T_m = 1.0$, while for $T/T_m > 1.0$,
there is an increase in the number of topological defects accompanied
by an increase in the ratchet efficiency.
The ratchet motion reaches its greatest value
near $T/T_{m} = 1.0$, above which it decreases again.
This suggests that a ratchet effect can
still occur in the Wigner liquid regime
provided that the substrate is sufficiently strong.

\subsection{Negative Diode Effect}

\begin{figure}
\includegraphics[width=\columnwidth]{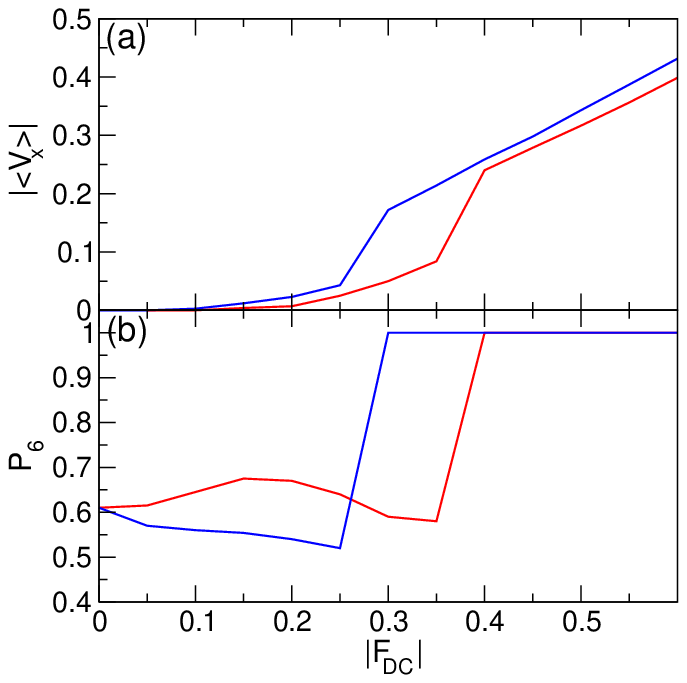}
\caption{
(a)  $|\langle V_{x}\rangle|$
and (b) $P_{6}$ vs $|F_{DC}|$
for a system with $\rho = 1.29$, $A_{p} = 1.25$,
and $N_{p}/L = 0.47$
under dc driving in the hard (blue) and easy (red) direction,
showing a negative diode effect.
There is also a transition
from a plastic to an elastic flow regime at higher drives.
}
\label{fig:20}
\end{figure}

\begin{figure}
\includegraphics[width=\columnwidth]{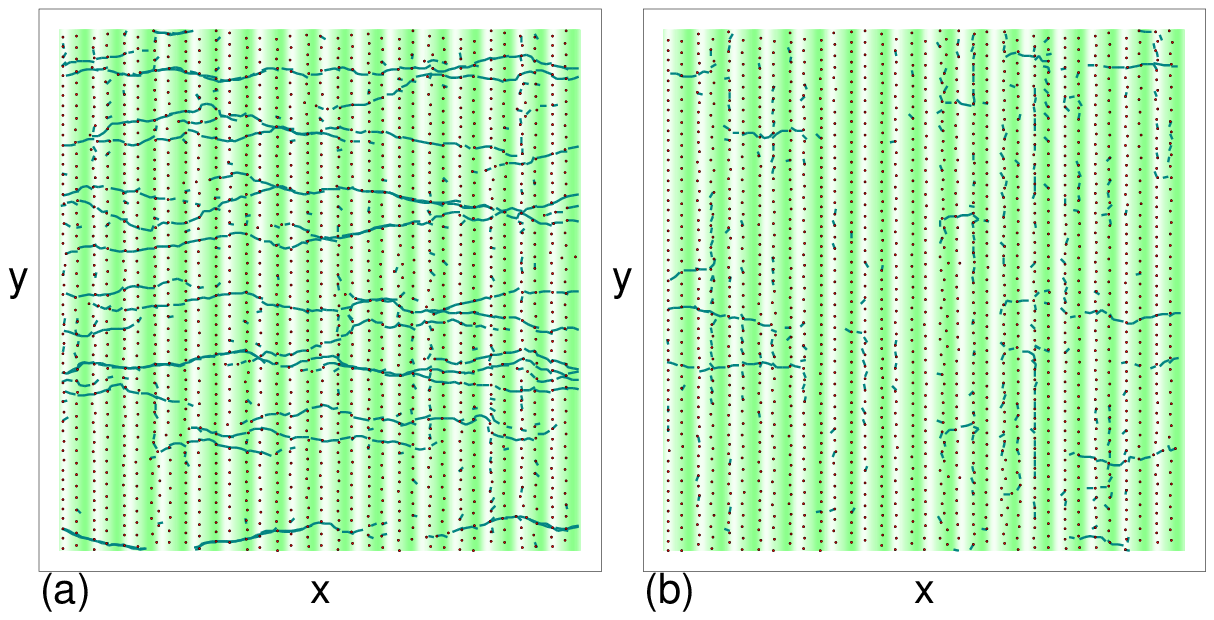}
\caption{
Images of the asymmetric substrate (green) and the particle trajectories
(blue) for the system in Fig.~\ref{fig:20}
with $\rho=1.29$, $A_p=1.25$, and $N_p/L=0.47$
under dc driving with $|F_{DC}| = 0.15$.
(a) Driving in the hard direction where there are plastic flow channels.
(b) Driving in the easy direction where the flow is suppressed.
}
\label{fig:21}
\end{figure}

We next show that when a negative ratchet effect occurs,
there can also be a negative diode effect
in which the flow is easiest along the hard direction
of the asymmetric substrate potential.
In Fig.~\ref{fig:20}(a),
we plot $|\langle V_{x}\rangle|$ versus $|F_{DC}|$ for
dc driving in the hard
and easy directions in a system with $A_{p} = 1.25$,
$\rho = 1.29$, and $N_{p}/L = 0.47$,
while Fig.~\ref{fig:20}(b) shows the corresponding
$P_{6}$ versus $|F_{DC}|$.
We find that
$|\langle V_{x}\rangle|$ is higher
for driving in the hard direction
than for driving in the easy direction,
and a jump up in
$|\langle V_{x}\rangle|$ coincides with a
transition from plastic to elastic flow.
Even within the elastic flow regime, the velocity is slightly
higher for driving in the hard direction.
For higher dc drives (not shown),
the velocity curves for the two driving directions become
almost the same.
In Fig.~\ref{fig:21}(a) we illustrate the particle
trajectories in the plastic flow phase at
$|F_{DC}| = 0.15$ for driving in the hard direction
where there are clear channels of flow,
while for the same value of $|F_{DC}|$ applied in the easy direction,
Fig.~\ref{fig:21}(b) shows that the
flow is almost zero. This indicates that the negative ratchet effect in
the plastic regime occurs via
the formation of localized excitations
that are more mobile.
Similar defects do not appear for driving in the easy direction
because the particle lattice can more
easily relieve the strain
induced by the drive since there are a larger number of
possible configurations that
the particles can adopt while still maximizing the pinning energy
when the drive is in the easy direction.
The negative ratchet effect is the most
pronounced at low drives in a regime where elastic flow can occur
for driving in the hard direction but
there is still plastic flow for driving in the easy direction.

\section{Discussion}

Our results suggest that
a variety of pronounced ratchet effects can occur in Wigner 
crystal systems provided that the spacing of the
substrate is within a few multiples of the
average spacing of the particles.
This ratchet effect should remain robust in a portion of the
Wigner liquid regime.
Other effects that we do not consider
here could also come into play, such as
the presence of random point defects that would destroy the lattice order.
In this
case, the commensuration effects would be reduced, but a ratchet effect would
still be possible.
If a magnetic field were applied to the system, the particles would move
with a finite Hall angle.
We have some preliminary data on the impact of a magnetic
field that will be the subject of a future work,
and we note that not only do ratchet effects still occur,
but new types of transverse ratchet motions are possible,
similar to those found for skyrmion ratchet systems
where Magnus forces are important \cite{Reichhardt15}.
As the electron
filling or density is varied,
it is possible for bubble, stripe, and electron liquid
crystal phases to occur in certain systems that
support Wigner crystals \cite{Fogler96,Lilly99,Gores07,Reichhardt10}.
It is not clear what would happen to the ratchet effect if there
were a transition from a Wigner crystal to a stripe or bubble phase.
Ratchet effects could still
be possible, but the necessary length scale of the substrate could change.
For example, substrates with
much larger periods than what we consider here
might produce more effective
ratcheting motion since the lattice constant of the bubble or stripe could
be larger than that of the Wigner crystal.
In addition to transport, other measures
that can be used to detect ratchet motion include
noise measurements or some resonance measures.
A particularly interesting aspect of our work is
the fact that the structure of the particle lattice is strongly
affected by the stage of the ac driving cycle.
In some cases, the particles have
strong triangular ordering for one direction of ac driving
and a more liquid structure for the other driving direction.
This suggests that other types of structural
measures or measures that couple to the long
range order of the system could be used
to detect ratcheting motion.
Our results
have implications for other kinds of collectively interacting
particle systems undergoing ratcheting
since we have observed
both elastic and plastic ratchet regimes as well
as mixed phases.
Many
of the structural rearrangements that we find under ac
driving will also occur in other systems
with long range or intermediate range
particle-particle interactions when coupled to this type of
asymmetric substrate, including charged colloids, dusty plasmas, and
superconducting vortices in thin films.

\section{Summary} 
We have investigated the diode and ratchet behavior for
a two-dimensional Wigner crystal coupled
to an asymmetric one-dimensional substrate.
We show that strong collective effects
give rise to a variety of different phases. The system
undergoes a single depinning transition for the
dc driven diode
motion found at low densities,
while in the collective regime, this depinning threshold
is strongly reduced.
At higher dc drives,
there is a two-step depinning process for motion in the hard
direction of the asymmetric substrate, along with
an extended window of drives in which the velocity for driving
in the hard direction is smaller than for driving in the easy direction.
For ac driving, we show that there are multiple
ratcheting regimes,
including an elastic ratchet where the particles keep the same neighbors
throughout the ac drive cycle,
a plastic ratchet where the system can be disordered for one or both halves
of the ac drive cycle,
a diode ratchet where the particles only hop
in one direction during an ac drive cycle, and a pinned
phase.
The maximum ratchet effect occurs at the crossover between the
plastic and diode ratchet regimes.
For weak substrates or higher fillings,
the particles can effectively decouple from the substrate in
an Aubry transition and the ratchet effect is lost.
In addition to the phases listed above,
we also find several commensurate phases for
certain fillings where the system forms an ordered
triangular, square, or distorted lattice.
These commensurate states show elastic diode ratchet effects and
can produce local maxima or minima in the ratchet efficiency.
We show that the
particle structure depends strongly on both the ratchet regime
and also the stage of the ac drive cycle.
In some cases, the flow is more ordered during the positive portion of the
ac drive cycle but becomes disordered during the
negative portion of the cycle.
We also find that at high fillings, the system can exhibit
a reversed ratchet where the particles move more easily
along the hard direction of the substrate asymmetry, which can
lead to the appearance of a negative diode effect.
Our results suggest that a new way to test for the presence of a Wigner crystal
is by looking for ratchet effects.
Our results also show that significant structural transitions
can occur during the course of the ac drive cycle in the collective regime,
which may also arise in other strongly coupled systems interacting with
asymmetric substrates.

\begin{acknowledgements}
We gratefully acknowledge the support of the U.S. Department of
Energy through the LANL/LDRD program for this work.
This work was supported by the US Department of Energy through
the Los Alamos National Laboratory.  Los Alamos National Laboratory is
operated by Triad National Security, LLC, for the National Nuclear Security
Administration of the U. S. Department of Energy (Contract No. 892333218NCA000001).
\end{acknowledgements}

\bibliography{mybib}

\end{document}